\documentclass[sigconf]{acmart}
\AtBeginDocument{%
 \providecommand\BibTeX{{%
  \normalfont B\kern-0.5em{\scshape i\kern-0.25em b}\kern-0.8em\TeX}}}

\copyrightyear{2023}
\acmYear{2023}
\setcopyright{acmlicensed}\acmConference[FAccT '23]{2023 ACM Conference on Fairness, Accountability, and Transparency}{June 12--15, 2023}{Chicago, IL, USA}
\acmBooktitle{2023 ACM Conference on Fairness, Accountability, and Transparency (FAccT '23), June 12--15, 2023, Chicago, IL, USA}
\acmPrice{15.00}
\acmDOI{10.1145/3593013.3594083}
\acmISBN{979-8-4007-0192-4/23/06}

\usepackage{hyperref}
 \usepackage[compact]{titlesec}
	  \titlespacing*{\section}{0pt}{1.8ex}{0.6ex}
	  \titlespacing*{\subsection}{0pt}{0.8ex}{0ex}

\begin{document}

\title[]{An Empirical Analysis of Racial Categories in the Algorithmic Fairness Literature}

\author{Amina A. Abdu}
\email{aabdu@umich.edu}
\affiliation{%
 \institution{University of Michigan}
 \streetaddress{}
 \city{Ann Arbor}
 \state{Michigan}
 \country{USA}
 \postcode{48104}
}

\author{Irene V. Pasquetto}
\email{irenevp@umich.edu}
\affiliation{%
 \institution{University of Michigan}
 \streetaddress{}
 \city{Ann Arbor}
 \state{Michigan}
 \country{USA}
 \postcode{48104}
}

\author{Abigail Z. Jacobs}
\email{azjacobs@umich.edu}
\affiliation{%
 \institution{University of Michigan}
 \streetaddress{}
 \city{Ann Arbor}
 \state{Michigan}
 \country{USA}
 \postcode{48104}
}


\begin{abstract}
 Recent work in algorithmic fairness has highlighted the challenge of defining racial categories for the purposes of anti-discrimination. These challenges are not new but have previously fallen to the state, which enacts race through government statistics, policies, and evidentiary standards in anti-discrimination law. Drawing on the history of state race-making, we examine how longstanding questions about the nature of race and discrimination appear within the algorithmic fairness literature. Through a content analysis of 60 papers published at FAccT between 2018 and 2020, we analyze how race is conceptualized and formalized in algorithmic fairness frameworks. We note that differing notions of race are adopted inconsistently, at times even within a single analysis. We also explore the institutional influences and values associated with these choices. While we find that categories used in algorithmic fairness work often echo legal frameworks, we demonstrate that values from academic computer science play an equally important role in the construction of racial categories. Finally, we examine the reasoning behind different operationalizations of race, finding that few papers explicitly describe their choices and even fewer justify them. We argue that the construction of racial categories is a value-laden process with significant social and political consequences for the project of algorithmic fairness. The widespread lack of justification around the operationalization of race reflects institutional norms that allow these political decisions to remain obscured within the backstage of knowledge production.
 
\end{abstract}



\keywords{racial categories, algorithmic fairness, state race-making}


\maketitle

\section{Introduction}
Historically, racial classification for the purpose of detecting discrimination has been codified in the U.S. context through the state via legislation and federal agencies like the Census Bureau \cite{morning05}. More recently, however, the field of algorithmic fairness has 
begun to act as a mechanism, often extra-legal, for 
identifying and preventing discrimination in sociotechnical systems. Thus decisions about how to define and formalize racial categories have been relocated from policymakers within government to academics and technologists involved in the design and implementation of algorithmic fairness. Within the algorithmic fairness community, a recent body of work has emerged around the adoption of formal racial categories \cite{benthall19, hanna20, kasirzadeh21, lu22}. This nascent literature challenges the uncritical adoption of legal and biological notions of race and calls on the algorithmic fairness community to explore interventions that appropriately account for the socially situated nature of race. In the absence of such interventions, algorithmic fairness researchers risk undermining their efforts toward anti-discrimination by reifying harmful social divisions \cite{benthall19}, failing to correctly measure racial disparities \cite{hanna20}, and proposing solutions that are misaligned with the problems they purport to solve \cite{roth10}. Mitigating these risks, however, requires a better understanding of current practices around the operationalization of race within the algorithmic fairness community. Through a systematic content analysis of papers published in the initial years of the ACM Conference on Fairness, Accountability, and Transparency (FAccT), this paper provides an empirical assessment of the construction of racial categories in the emerging algorithmic fairness community. We examine these classification practices within the context of the longstanding history of state and scientific institutions in race-making. Through this analysis, we aim to identify how algorithmic fairness researchers operationalize and conceptualize race and how they justify their decisions around the operationalization of race in order to understand the institutional influences that shape racial classification practices within the academic algorithmic fairness community. 

 Drawing on the history of institutional race-making, we perform a qualitative analysis of papers in the algorithmic fairness literature that engage with issues of racial discrimination in sociotechnical systems. Specifically, we conduct hybrid inductive-deductive content analysis on a sample of 60 papers published at FAccT between 2018 and 2020 that focus on fairness issues and reference race. Using a combination of structured and thematic analysis, we identify schemas used for operationalizing race in general and multiracialism in particular. We examine various characteristics-- including legal protection, social status, appearance, and identity-- through which group differences are constructed in the algorithmic fairness literature. We also examine the justification that researchers provide or fail to provide when adopting a particular conception of race. Finally, we explore the values associated with these choices. 

While we find that categories used in algorithmic fairness work often echo legal frameworks, as has been proposed in the existing critical literature \cite{benthall19, hanna20}, we argue that values from theoretical computer science play an equally important role in the construction of racial categories. These influences must be understood in order to assess the extent to which racial classification under algorithmic fairness frameworks departs from other institutional understandings of race. This shift has important implications for how the values and institutions in the algorithmic fairness community have shaped practices of racial classification. Moreover, misalignment between legal frameworks and algorithmic fairness frameworks has consequences for the utility and impact of algorithmic fairness interventions in real-world settings.

This work serves as a case study for understanding how normative values are adopted, embedded within, and obscured by analytic choices about how to measure, quantify, and represent social categories. In surfacing the relationship between values and analytic choices, this project represents the beginning of a research agenda to ensure that algorithmic fairness research is in fact working toward its intended anti-discrimination goals rather than uncritically reproducing existing power relations. \looseness=-1

\section{Related Work} 

\subsection{Racial Categories in Algorithmic Fairness}
The literature on racial classification in algorithmic fairness frameworks highlights a lack of attention toward the nature of racial categories. Much work has been done in computer science to formally define fairness, leading to significant work surrounding the conflict between notions of group and individual fairness and understanding what is meant by ``fairness.'' However, there has been less work around what is meant by ``group'' \cite{benthall19}. Critiques of algorithmic fairness frameworks highlight the mistreatment of race as an individual trait rather than relational system \cite{kohler18, benthall19}, insufficient attention to the situated and context-dependent nature of race \cite{hanna20, kasirzadeh21, lu22}, and the uncritical adoption of the ``protected class'' framework of race from U.S. anti-discrimination law \cite{benthall19, hanna20}. Moreover, this body of work argues that by failing to engage meaningfully with the meaning of social categories, algorithmic fairness frameworks are susceptible to adopting incoherent and dangerous notions of race that reduce racial distinctions to differences in biology or appearance \cite{kohler18, kasirzadeh21}. \looseness=-1

The literature in this area reveals that this problem is not unique to group fairness. While some critiques focus primarily on the failures of group fairness to account for differences between groups, critiques of counterfactual fairness--the most popular formalization of individual fairness \cite{kusner17}-- highlight the persistent problem of defining relevant categories of analysis \cite{kohler18, kasirzadeh21}. The counterfactual model of fairness proposes that a predictor is fair toward an individual if it would have given the same prediction in the counterfactual world where the individual had belonged to a different group, for example a racial group. Operationalizing this model of fairness requires confronting both what makes a counterfactual world similar enough for comparison and what it means for an individual to belong to a different racial group. Criticisms of the counterfactual model’s treatment of race demonstrate that computer scientists cannot escape the thorny political work of racial classification by using a particular mathematical definition of fairness, even one that purports to center individual merit over group membership.
\looseness=-1

\subsection{Identifying Values in Machine Learning Research} \label{values}

In recent years, there has been growing interest around specifying and uncovering the values embedded in machine learning research. Researchers have highlighted the importance of such values in shaping seemingly technical decisions \cite{chasalow21, hancox21, jacobs21}. In order to understand how normative values are embedded within decisions about how to operationalize race, we examine the values underlying algorithmic fairness research. Ethical considerations in technical research and AI include autonomy, beneficence, non-maleficence, justice, explicability, and legal compliance \cite{menlo, floridi21}. However, in practice, machine learning research tends to under-emphasize these ethical principles in favor of values like performance and efficiency \cite{scheuerman21, birhane22}. While the FAccT community explicitly centers the ethical values of fairness, accountability, and transparency, it often overlooks other moral values such as respect and agency \cite{laufer22}. \looseness=-1

Prior work on values in machine learning research highlights the community's tendency to prioritize generalization, universality, and abstraction over values of contextuality and situatedness \cite{hancox21, scheuerman21, birhane22}. This pattern exists in both the broad machine learning community and within the algorithmic fairness community in particular, where researchers often fail to name concrete harms and specific impacted groups, for example failing to directly address anti-Blackness \cite{birhane22_2}. 
The literature highlights two fundamental tensions: the tension between ethical and performance values \cite{birhane22} and the tension between generalizability and contextuality \cite{scheuerman21}. We focus on these key values to assess how values influence the adoption of racial categories. \looseness=-1

\section{Racial Classification and Institutions}

Institutions play a critical role in race-making; science and the state have been particularly influential sites in the creation and designation of racial categories. 
We propose that the algorithmic fairness community is an emerging race-making institution that merits further attention. Although prior work has primarily highlighted the dangers of algorithmic fairness researchers uncritically reproducing legal and biological conceptualizations of race, we argue that it is equally important to understand how algorithmic fairness frameworks align with and depart from these traditional institutional influences. In particular, we emphasize that the algorithmic fairness community has its own values, goals, and practices that shape the adoption and construction of racial categories, which we explore in our analysis.
For greater context, we first present an incomplete overview of this history to demonstrate how institutional contexts, values, and goals have shaped racial classification practices.

\subsection{Racial Classification in Scientific Research}

The scientific enterprise engages in classification by identifying kinds of people \cite{hacking06}, which serves as an important site of political and ethical work \cite{bowker00}. 
Dorothy Roberts argues that modern racial classifications emerged jointly from the scientific revolution and colonialist expansion to create and bolster new state and scientific institutions \cite{roberts2011fatal}.
Race became a project of biological classification---whether to evidence or establish a scientific basis of racial differences (e.g., Linnaeus, Galton) or undermine it (Darwin)---that remains a foundation of scientific, social, and medical research \cite{roberts2011fatal,vyas2020hidden}. In the U.S., projects of governance (the census, voting, citizenship) and trade and political projects (from slavery and abolition \cite{rosenthal2019accounting} to eugenics \cite{roberts2011fatal,porter2018genetics}) formed a route for scientific racism to become encoded in social projects \cite{roberts2011fatal}.
\looseness=-1

On one hand we might observe social and cultural nuance: Roth \cite{roth12} theorizes \textit{racial schemas} as cognitive and cultural classification processes that can vary from person to person, even acknowledging that one person can hold multiple racial schemas at once. Despite variation, the construction of boundaries between groups is shaped by a variety of political and social factors including institutions, power, and political network structures \cite{wimmer08}.
Yet within modern practices of science, we see the history of race as a governing technology play out today \cite{roberts2011fatal}. 
\looseness=-1
For instance, when researchers use racial categories in their studies, race is frequently conceptualized as a fixed, and often biological, identity characteristic rather than a dynamic social and political phenomenon \cite{zuberi00, james08}. \looseness=-1
Scientists may choose a given racial classification for a variety of reasons, including widespread acceptance, the ability to facilitate comparisons across studies, and stability \cite{smart08}. Inconsistencies in racial categories have been noted in many disciplines including survey methods \cite{saperstein06}, public health \cite{kaufman99, magana17}, and computer vision \cite{scheuerman20, khan21}. Although differences in racial classification can affect research conclusions \cite{saperstein06}, researchers often fail to explain or justify their operationalizations of race \cite{lee09, scheuerman20}. This has the potential to reify harmful conceptions of race and undermine the effectiveness of interventions intended to address racial disparities \cite{lett22,scheuerman2021auto} \looseness=-1
and instead obscures that fundamentally arbitrary, inconsistent racial classifications are ideological and political \cite{omi14}.

\subsection{State Race-Making} \label{racemaking}

The state plays an essential role in making and enforcing racial categories through censuses, legislation, and everyday governance \cite{brown20}. These categories serve as powerful tools for social stratification and reflect normative decisions about how states ought to allocate resources and rights. Brown \cite{brown20} identifies three institutional characteristics in particular which shape state racial classification: evidentiary standards for decision-making, record-keeping requirements, and incentive structures. We return to the role of these three institutional structures in algorithmic fairness in the discussion.
\looseness=-1

In their work on Indigenous statistics, Walter and Andersen \cite{walter13} draw an important link between the creation of such racial categories and quantification, noting that the statistical representation of Indigeneity is an explicit project of racialization. They highlight the role of power, and particularly of state power, in the formation of racial categories through data collection and statistical analysis. While Walter and Andersen focus on official population statistics, such as censuses, they note that quantification extends beyond this particular setting. Indeed, quantitative representations of race are central to the project of algorithmic fairness. Building on Walter and Andersen’s analysis of state power, we propose that the algorithmic fairness community acts as an emerging site of power through its quantitative enactment of racial boundaries. 
\looseness=-1

\subsection{Institutional Goals in Classification: The Case of Multiracial Identity} \label{stategoals}
The field of critical mixed race studies provides a framework for engaging with historical and contemporary state efforts to construct race toward its own ends. 
We briefly discuss two examples where developments in the state project of race-making, towards ostensibly inclusive ends, were used to reinforce the dominant hierarchy.
In the U.S. context, the political meaning of multiracialism has evolved from the legacy of the one-drop rule---a racial classification principle that asserts that a person with any Black ancestry should be classified as Black---to a celebration of multiracialism as emblematic of a post-racial American future during the introduction of multiracial identification on the 2000 census. This history brings to light several political values embedded within emergent conceptions of multiracialism. Multiracialism came to be depicted as an antidote to historical racial divides, closely linked to American national identity and the image of the U.S. as a ``melting pot" of cultures. Moreover, multiracialism was associated with the future, positioning Black identity politics as dated and reinforcing the logic of white supremacy by creating a new economically, politically, and socially ascendant racial identity through its distance from Blackness \cite{essi17}.
This trajectory is not unique to the U.S. context. In post-revolutionary Mexico, a new \textit{mestizo} identity was forged around modernity and nationalism \cite{dalton18}. Following the Mexican Revolution, the Mexican middle class--which was primarily Indigenous--gained social and economic power. Rather than respond to their interests, however, the Mexican government formed a hybrid racial identity whose modern goals would align with the state’s goals of industrialization and economic development. Moreover, implicit in this new identity was a distance from Indigeneity, which could be escaped through the adoption of technology and assimilation to the new mestizo identity.
The historical construction of multiracial identity across both the U.S. and Mexican contexts demonstrates the political goals embedded within the decision of how multiracial individuals are racialized. In each case, new classification systems were ultimately used to reproduce and reinforce the structure of the dominant racial hierarchy.

\section{Method}
In order to empirically assess the construction of racial categories in the algorithmic fairness community, we performed qualitative coding on a set of papers from the algorithmic fairness literature that discuss race between 2018 and 2020. While not exhaustive of the entirety of the fairness community, this allows us to identify and discuss emerging notions of race within such a community in its nascent years. Given our limited sample and the qualitative nature of our research, we have no claims over the generalizability of our findings beyond our sample, but, because of the criteria we used for selecting and analyzing our sample, we believe that the set is nevertheless a fair and telling snapshot of how race is conceptualized in recent literature on algorithmic fairness. The following subsections describe the details of the sample construction and the coding process. Qualitative coding of papers enables us to focus on realized research practices within the algorithmic fairness community. Future work might draw on interviews with researchers to examine how authors' perspectives and intentions interact with these practices, but this is beyond the scope of our work.

\subsection{Sample}
Data collection and analysis was performed by the first author of this paper. The author constructed the sample by beginning with every paper published in a leading domain-specific conference within the algorithmic fairness community, the ACM conference on Fairness, Accountability, and Transparency (FAccT, originally FAT*), between 2018-2020. As a flagship publication venue for work on algorithmic fairness, FAccT sets the standards for work in this area and broadly represents the state of the field. The first three years of the conference were examined to understand the process by which categories emerge and become naturalized within a nascent community of practice. To target works primarily about algorithmic fairness, the author adapted the selection criteria proposed in Fabris et al.’s \cite{fabris22} survey of data sets used in the algorithmic fairness literature and selected the subsample of these papers whose abstract contains at least one of the following strings, where the asterisk represents the wildcard character: *fair* (targeting, for example ``fairness'', ``unfair''), *bias* (``biased'', ``debias''), *discriminat* (``anti-discrimination'', ``discriminatory''), disparate, *parit* (``parity'', ``disparities''). From this subsample, only the papers that deal directly with race by restricting to papers which contain at least one of the following strings race, *racism (``racism", ``antiracism''), racial were selected. Finally, a manual check of this set of papers was performed and any papers that use these keywords with a different meaning were removed. This left a subset of 65 papers, of which 5 were extended abstracts rather than full papers. The extended abstracts were excluded from analysis because space constraints significantly limit the extent to which authors can explain and justify their choice of race categories. Thus, the final sample comprised 60 full-length FAccT papers published between 2018 and 2020.

\subsection{Analysis}
To analyze the sample of papers, systematic qualitative coding was performed via ATLAS.ti. Following a semi-grounded theory approach, the author employed an iterative qualitative coding process including an initial coding stage, which draws upon the theoretical literature outlined above, and a subsequent focused coding stage where the author reorganized and synthesized the data coded in the initial stage in order to identify and verify emergent patterns. In the initial coding stage, each paper was coded line-by-line for conceptualizations of race and values. Initially, the author searched for conceptualizations of race emphasized in the existing literature (for example, legal constructions of race). When alternative ways of conceptualizing race appeared, these were coded using an in vivo coding approach, which emphasizes preserving the exact terms used in the text. A similar process was used for coding values in the annotated documents: the author initially drew on existing ethical frameworks for AI and prior work on values in machine learning described in section \ref{values}, focusing in particular on ethical and performance values, generalizability, and contextuality. Other values were added through in vivo coding. Following this open coding stage, the author used the initial list of codes to go through each document again to ensure that codes that emerged mid-process were applied to each paper. At this point, focused coding was employed to identify the most frequent and significant codes \cite{saldana16} and to compare codes to understand connections and themes. Based on this, closely related codes were grouped into broader categories (for example, values of efficiency, effectiveness, accuracy, and performance were all grouped under performance-related values). Reflections on these choices were documented in analytic memos.

Line-by-line coding was supplemented with structured paper-level codes about each document in the corpus. Specifically for each paper, the author noted operationalizations of race, operationalizations of multiracialism, and direct quotes of any justification provided for the choice of schema.

After coding each paper at a line-by-line level according to \textit{conceptualizations} of race and \textit{values}, codes were summarized at the paper level in a matrix display alongside the \textit{justifications} collected in the paper-level codes. The author used the matrix display as described by Miles et al.\ \cite{miles18} to visualize a condensed version of the data for verification and analysis. The side-by-side visualization of these three constructs enabled us to reflect on their collective meaning under the assumption that they form an interconnected system through which racial categories are adopted. In the sections that follow, we highlight the frequency of relevant concepts and draw on textual evidence to surface both what is present and what is absent in the algorithmic fairness literature's treatment of race. We discuss operationalizations and conceptualizations of race, researcher justifications for these choices, and the values associated with these choices.

\section{Construction of Racial Categories in FA\MakeLowercase{cc}T Papers, 2018-2020} 
We begin by examining how racial categories are formalized in the annotated documents. 
We then discuss the specific case of multiracialism to illuminate broader patterns in how race is operationalized. Finally, we explore how boundaries between racial groups are conceptualized more generally. We highlight inconsistencies in both the formalization and conceptualization of race, reflecting diverging understandings of why race is an important site of study and of intervention within the algorithmic fairness community.
\looseness=-1

\begin{figure}[h]
 \centering
 \includegraphics[width=\linewidth]{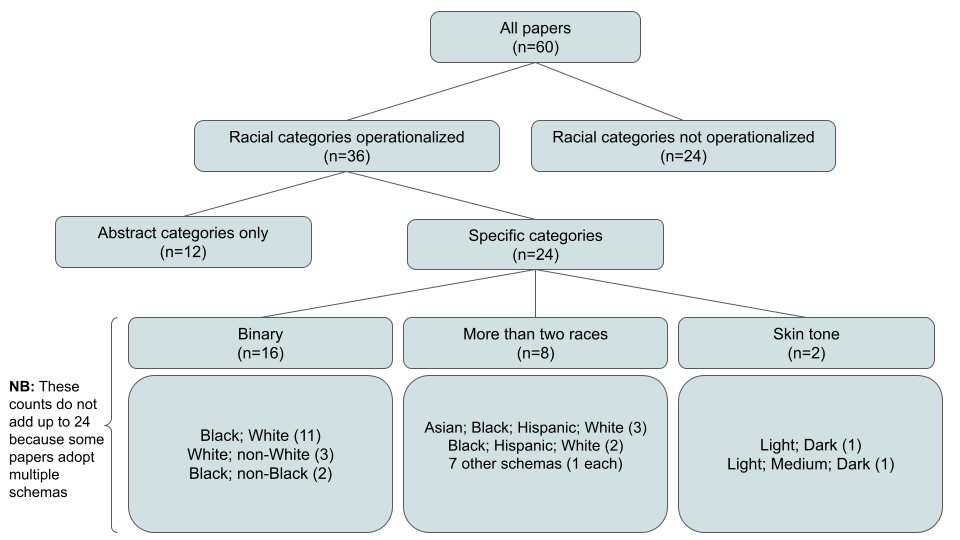}
 \caption{An overview of how race is formalized within the annotated papers. Counts of papers that adopt each formalization are included in parentheses. The 24 papers that do not use racial categories, i.e., the top right leaf, mention race as an example, but do not return to it, or discuss race at a purely conceptual level and do not formalize racial categories.}
 \label{fig: schemas-overview}
 \Description{A schematic overview of the ways that race is formalized in the annotated papers.}
\end{figure}

\subsection{Racial categories are adopted inconsistently} 
There is significant inconsistency in the way that racial categories are adopted within the algorithmic fairness literature. We find no evidence that algorithmic fairness researchers defer to common norms or institutional standards for defining race (for example, Census categories). The categories used vary from paper to paper, and sometimes even within a single paper. Race is conceptualized at varying levels of abstraction and different sets of categories are deemed relevant for analysis. 
Figure \ref{fig: schemas-overview} shows an overview; additional details can be found in Table~\ref{table:race_category_types} in the Appendix.
\looseness=-1

36 of the 60 FAccT papers analyzed provide some formal notion of racial categories.
Of these, 12 (33.3\%) leave the racial categories in question abstract, to be defined at the algorithm implementation stage. Typically, these abstractions not only allow for multiple racial classification schemas but also enable the substitution of any group in place of a racial group. Under this model, racial categories are seen as interchangeable with other social categories, legally protected groups, or groups of different sizes.
\looseness=-1

Among the 24 papers that define specific race categories, there are 14 distinct racial classification schemas used, which fall into 5 broad categories: Black/white, white/non-white, Black/non-Black, more than two races, and skin tone. The most common of these, used in 11 papers, is a binary classification schema that distinguishes between white and Black or closely related categories like Caucasian and African-American. Less commonly used are binary schemas that distinguish between white and non-white (3 papers) or Black and non-Black (2 papers). Eight papers adopted classification schemas with more than two race categories. Among these 8 papers, there were 9 different categorization schemas, indicating little agreement about what categories are relevant for analysis (\textit{N.B.}\ some papers adopt multiple schemas). All of these schemas include Black and white categories, demonstrating a shared understanding of these categories' importance. Asian and Hispanic groups appeared in 5 and 7 of these schemas respectively indicating some agreement about their relevance.
Finally, two papers, both in the computer vision setting, used a measure of skin tone as a proxy for race rather than directly adopting racial categories. 
\looseness=-1

Decisions about how to delimit racial categories vary not only between papers but sometimes within a single paper. Four papers adopt multiple schemas for race, reporting different results using different sets of racial categories. In all but one case, the schemas in a given paper were subsets of one another. For example, one paper presents results using the following three classification schemas: 1.\ Asian, Black, Hispanic, Mixed, White, Other, Unknown; 2.\ Black, Mixed, White, Other, Unknown; and 3.\ Black, White. Typically the use of multiple schemas is not explained. In the previous example, for instance, it is unclear both whether the ``Other'' category is modified to include the Asian and Hispanic categories under Schema 2 and why the schemas differed between analyses. Only one paper presents a justification for using multiple classification schemas, citing statistical robustness as a reason to collapse groups with a small number of observations under a single umbrella.
\looseness=-1

These findings highlight a wide variety of racial classification schemas present across the literature. While there is not a consensus or standard view of the full set of racial categories that are relevant in algorithmic fairness research, there is widespread agreement about the importance of the Black and white categories. Interestingly, despite the historical and institutional importance of the Census Bureau in defining racial categories, no paper used the exact schema used to collect race data in the decennial Census. Rather than deferring to historically influential standards, current norms within the algorithmic fairness community grant researchers substantial autonomy in their ability to select a racial classification schema.
\looseness=-1

\subsection{Multiracialism is often elided} 

Multiracialism is rarely mentioned in the annotated papers. Of the 24 papers that use specific race categories, only 3 (12.5\%) disclose how multiracial people are classified. Each of these papers uses a different process: one combines all multiracial observations into a single ``mixed'' label, one includes multiracial observations under the larger umbrella of ``other'', and one excludes multiracial observations from analysis.
\looseness=-1

The lack of a multiracial category in all but one paper reflects the tendency of papers within the literature to adopt binary classification schemas. However, even within these schemas it remains unclear how multiracial people are classified. In the frequently adopted Black/white schema, for example, it is unclear how data points representing multiracial Black and multiracial white individuals are treated. This decision enables significant analytic flexibility on the part of the researcher between several justifiable options. Researchers may choose to exclude multiracial observations from analysis; to count them with each group of which they are a member (for example, including a biracial Black and white person in analyses of both the Black population and the white population); or to count them only within the historically marginalized group, among other options. As the multiracial population grows, these decisions will increasingly influence the results of fairness analyses. While our findings show that current publishing norms allow these decisions to remain concealed, this obscurity may enable the manipulation of the multiracial category toward hidden ends.
\looseness=-1

\subsection{Group boundaries are constructed across many dimensions} \label{groupboundaries}
The inconsistency in how researchers construct racial categories reflects deeper inconsistencies in how the algorithmic fairness community understands race. Differences between racial groups are conceptualized in a number of ways both across and within papers. Underpinning these inconsistencies are divergent views of what types of categories are relevant for analysis. We discuss the five most common conceptualizations of racial difference below: legal protection (45\% of papers), social status, minority status (28.3\%), sensitivity (28.3\%), and social salience (16.7\% of papers). 
For a full breakdown of all conceptualizations, including counts and example quotes from the data, see Table~\ref{table:noname_types} in the Appendix. \looseness=-1

\subsubsection{Legal Protection.\ }
Prior work on race in algorithmic fairness frameworks highlights the prevalence of conceptualizing racial groups as ``protected classes'' and establishing group boundaries in terms of legal protection \cite{benthall19, hanna20}. Indeed, this was the most common way of describing race in the literature, appearing in 27 (45\%) of the annotated papers. As in the following example, recourse to protected classes often refers to U.S. law and views race as interchangeable with other legally protected attributes (in particular, gender):
\begin{quote}
  ``We consider \textbf{U.S. anti-discrimination laws}, which name \textbf{race, color, national origin, religion, sex, gender, sexual orientation, disability, age, military history, and family status} as protected attributes'' -- Yang et al.\ 2020 (p. 553)
\end{quote}
Legal frameworks may be particularly useful for aligning algorithmic fairness interventions with existing anti-discrimination law, but are also adopted in papers that do not explicitly attempt to support legal interventions. Despite its prevalence, this framework appears in fewer than half of the annotated papers and is far from the only conceptualization of race in the literature.

\subsubsection{Social Status.\ }
Notions of race that emphasize status 
distinguish between 
advantaged groups and disadvantaged groups as the relevant populations for comparison. Advantage or disadvantage is understood in a number of ways, including power, privilege, vulnerability, and stigma. Yet, under this vision of race, algorithmic fairness interventions attempt to mitigate the very disadvantage they use to distinguish between groups.
\looseness=-1

\subsubsection{Minority Status.\ }
Although minority status is often closely linked to processes of disadvantage, 
the conceptualization of racial groups in algorithmic fairness as minority and majority groups emphasizes a \emph{quantitative} understanding of group boundaries rather than one situated in political and social relations. Distinguishing between minority and majority groups may be particularly relevant in algorithmic fairness settings where unfairness arises in part because of the under-representation of minority groups within the data.
\looseness=-1

\subsubsection{Sensitivity.\ }
The term ``sensitive attribute'' is commonly used within the theoretical computer science literature on privacy, where it refers to information that would be undesirable to disclose. However, it is unclear what makes a given attribute sensitive when this term is adopted outside of the privacy context. Consequently, sensitivity represents a flexible way to talk about group differences. While the annotated documents rarely address the meaning of sensitivity, sensitivity is occasionally defined in terms of other notions of race. In the following quotes, for example, sensitivity is aligned with legal protection, social status (in this case expressed as privilege), and minority status:
\looseness=-1
\begin{quote}
  ``Historical datasets often reflect historical prejudices; \textbf{sensitive or protected attributes} may affect the observed treatments and outcomes.'' --Madras et al.\ 2019 (p.\ 349)
\end{quote}

\begin{quote}
  ``$S$ is the sensitive attribute where [$S$ =1] is the \textbf{privileged class}.'' --Friedler et al.\ 2019 (p.\ 332)
\end{quote}

\begin{quote}
  ``An intentionally malicious—or unintentionally ignorant— advertiser could leverage such data to preferentially target (i.e., include or exclude from targeting) users belonging to certain sensitive social groups (e.g., \textbf{minority race}, religion, or sexual orientation).'' --Speicher et al.\ 2018 (p.\ 2)
\end{quote}
Because of the ambiguous meaning of sensitivity, the sensitive attribute terminology provides an abstract and general way to discuss groups without engaging with the meaning of these groups.

\subsubsection{Social Salience.\ }
 The critical literature on racial categories in the algorithmic fairness literature emphasizes the notion that racial categories should refer to ``socially salient'' groups \cite{benthall19, hanna20}. In other words, racial groups should be studied according to the relevance of their group membership in a particular social context. This perspective has been adopted occasionally in empirical settings: in the following quote, for example, the authors indicate that a result should be ignored because the group in question is merely an artifact of the data and does not describe a relevant social group out in the world:
 \looseness=-1
\begin{quote}
  ``the most significant difference—that between the ``Unknown'' category and the rest— is \textbf{not one that directly corresponds to a salient race/ethnicity group}'' --Chouldechova et al.\ 2018 (p. 9)
\end{quote}

\subsubsection{Uncommon Conceptualizations of Race.\ }
In addition to the conceptualizations of race that appear frequently, it is worth noting conceptualizations that remain largely absent. While numerous ways of conceptualizing race appear in the algorithmic fairness literature, we note that race is typically treated as a legal, social, or political axis of discrimination rather than an issue of personal identity. Only about 6.7\% of papers conceptualize race according to personal identity. This aligns with the FAccT community's general orientation toward values of fairness and justice over values like dignity, respect, and self-determination \cite{laufer22}. We also note that biological notions of race appear infrequently: 10\% of papers highlight observable differences and only 3.3\% of papers discuss ethnic origin.
 \looseness=-1

\section{Researcher Justifications} 
We find that researchers typically fail to justify their choice of a particular racial categorization schema. In total, only 13 papers (21.7\%) provide any reasoning behind their chosen schema. Even when subsetting to the 24 papers that adopt a specific categorization schema, only 9 (37.5\%) present some form of justification. When justifications are provided, they fall into five broad categories: data availability, technical factors, appeals to prior work, epistemic concerns, and relevance. These types of justifications are not mutually exclusive, and are often inter-related. We summarize each type of justification below, with relevant examples from the annotated papers. (See Table~\ref{table:justification_types} in Appendix for all counts.) 
 \looseness=-1
 
\subsection{Data Availability}
Researchers may adopt a particular racial categorization schema based on how race is presented in the data they use for analysis. Researchers may choose not to modify the schema used in their data for analytic simplicity, as in the following example:
 \looseness=-1
\begin{quote}
  ``We use race and ethnicity as a combined field in this paper because \textbf{that is how the data was collected and organized} in the LA City Attorney’s Office system.'' --Rodolfa et al.\ 2020 (p.\ 147)
\end{quote}
In this case, the researchers default to the decision made by the Los Angeles City Attorney's Office for its own administrative purposes. Even if researchers choose not to adopt the same schema as in their data, they may still be affected by the choices of data collectors. To the extent that researchers use data that they did not originally collect, they may be limited by choices made at an earlier stage by someone else if relevant 
information is obscured under the original data collection schema. 
In the following example, race data are not collected at all, leading the researchers to use an arbitrary variable in its place:
\begin{quote}
  ``the first binary feature was used as a substitute sensitive feature since we \textbf{did not have access to sensitive features}.'' --Dwork et al.\ 2018 (p.\ 3)
\end{quote}
Finally, distrust in data quality may lead researchers to choose a particular schema: 
\begin{quote}
  ``\textbf{based on our analysis of the consistency of racial classification within the court data}, we have determined this categorization scheme introduces the fewest problems with inconsistent classification'' -- Lum et al.\ 2020 (p.\ 488)
\end{quote}
Based on these examples, we argue that data collectors exert influence over the racial categories adopted in algorithmic fairness both directly (by foreclosing certain analyses) and indirectly (through defaults and varying data quality).
\looseness=-1

\subsection{Technical Factors}
A number of technical considerations may influence a researcher's choice of racial categories. 
Algorithms used to identify or mitigate unfairness may be constrained in what types of inputs they can handle, particularly in the case of novel methods. This often leads to an emphasis on binary racial categorization schemas, such as the privileged/not-privileged dichotomy chosen in the example below:

\begin{quote}
  ``Some \textbf{algorithms additionally require that the sensitive attributes be binary} (e.g., ``White'' and ``not White'' instead of handling multiple racial categorizations) - for this version of the data (numerical+binary) we modify the given privileged group to be 1 and all other values to be 0.'' --Friedler et al.\ 2019 (p.\ 332)
\end{quote}

Computational efficiency for complex algorithms or in the analysis of large data sets may also lead researchers to choose simpler categorization schemas. In the following example, the researchers once again choose a two-category schema, this time distinguishing between Black and white.

\begin{quote}
  ``To make brute force auditing \textbf{computationally tractable}, we designate only two attributes as protected; \textit{pctwhite} and \textit{pctblack}, the percentage of each community that consists of white and black people respectively.'' --Kearns et al.\ 2019 (pp. 106-108)
\end{quote}

Finally, statistical robustness may motivate researchers to choose racial categorization schemas such that each category has a sufficiently large sample size. This could lead researchers to omit groups with small populations or to combine these groups into larger categories, as in the following example:
\begin{quote}
  ``[S]everal demographic categories appeared rarely, if at all, in the Twitter data. For the sake of \textbf{more robust statistical comparisons}, some analyses below collapse these race categories to, for example, \{\textit{White; Black; Hispanic; Other; Don’t Know}\}.'' -- Borradaile et al.\ 2020 (p. 574)
\end{quote}

In each case, technical constraints and desiderata lead researchers toward simpler categorization schemas with fewer racial categories. Though justifications for racial schemas are rare in the annotated papers, the prevalence of binary schemas suggests that technical motivations may play an important role in the adoption of racial categories within the algorithmic fairness community.

\subsection{Appeals to Prior Scientific Work}

Some justifications draw on prior academic research. These justifications often draw from beyond the algorithmic fairness literature, which is relatively new and has fewer established standards compared to, for example, the dermatology community cited in the following case.
\begin{quote}
  ``We chose the Fitzpatrick six-point labeling system to determine skin type labels given its \textbf{scientific origins}. Dermatologists use this scale as the \textbf{gold standard} for skin classification and determining risk for skin cancer'' -- Buolamwini \& Gebru 2018 (p.\ 6)
\end{quote}
However, as the algorithmic fairness community begins to establish its own norms, publication standards, and notions of rigor, future work in the field may instead appeal to existing work from within the community. In the following example, the authors cite the Buolamwini \& Gebru paper quoted above as justification for adopting a similar racial categorization schema in a similar context.
\begin{quote}
  ``\textbf{Similar to prior work}, skin color is used as a surrogate for race membership because it is more visually salient.'' --Yang et al.\ 2020 (p.\ 554)
\end{quote}
This process of self-perpetuation and naturalization points to the importance of norms and standards within algorithmic fairness institutions.

\subsection{Epistemic Concerns}
In addition to referencing specific scientific work, justifications of racial categorization schemas also draw on more general notions of scientific rigor by appealing to epistemic principles like reliability, consistency, objectivity, and precision: \looseness=-1
\begin{quote}
  `` Importantly, we determined that different coders following this protocol could \textbf{reliably classify the race and gender of users}. '' -- Borradaile et al.\ 2020 (p.\ 574) \looseness=-1
\end{quote}
\begin{quote}
  ``Since race and ethnic labels are unstable, we decided to use skin type as a more \textbf{visually precise} label to measure dataset diversity. Skin type is one phenotypic attribute that can be used to more \textbf{objectively characterize datasets} along with eye and nose shapes.'' -- Buolamwini \& Gebru 2018 (p.\ 4)
\end{quote}
As the algorithmic fairness community begins to establish its core epistemic values through publishing standards and methodological norms, these values will likely influence how researchers adopt racial categories and justify their choices.\looseness=-1

\subsection{Contextual Relevance}

Some papers justify their use of a particular categorization schema based on its relevance to the context of study. Researchers may adopt racial categories that reflect the cultural context in which the work is situated. In the case of the algorithmic fairness literature, researchers often draw on the U.S. context. As a result racial categories typically reflect notions of race stemming from the U.S.'s particular histories of slavery, segregation, and discriminatory policy. In the following justification, the researchers explicitly attempt to capture social understanding of race in the U.S. setting:
\begin{quote}
  `` Gender and race are fluid and socially constructed categories, and there are other possible ways of categorizing the gender and race of users. However, we believe these categories provide a reasonable, though necessarily simplified, \textbf{reflection of race and gender divisions in the US}.'' -- Borradaile et al.\ 2020 (p.\ 574) \looseness=-1
\end{quote}
Though justifications are rarely given in the annotated documents, cultural context can explain the prevalence of categorization schemas that center Blackness and whiteness. These categories of analysis are particularly relevant due to legacies of anti-Black racism and white supremacy. Beyond the larger cultural setting, justifications may also focus on racial categories' relevance in a particular domain of study. 
\begin{quote}
  ``Furthermore, skin type was chosen as a phenotypic \textbf{factor of interest} because default camera settings are calibrated to expose lighter-skinned individuals.'' -- Buolamwini \& Gebru 2018 (p.\ 4)
\end{quote}
The contextual relevance approach to racial categorization highlights the fact that inconsistencies across or even within papers are not necessarily a problem. Differences in racial schemas may reflect important differences in the social groups that are relevant to understanding and intervening in discrimination.

\section{Values in Classification} 
Racial classification is value-laden and political. Drawing on previous work establishing the values in machine learning and algorithmic fairness research, we identify the values that appear in the annotated documents in order to understand how normative goals drive the adoption of particular racial schemas. We find that the most frequently occurring values are performance-related (50\% of papers), which encompasses accuracy, effectiveness, and efficiency. This is followed by justice (45\%), which covers values like equity, equality, and merit; non-maleficence (36.7\%), which encompasses harm-reduction, risk-reduction, privacy, and safety; real-world applicability (33.3\%); epistemic values (28.3\%), which includes certainty, consistency, objectivity, and precision; contextuality (20\%); and generalizability (20\%). \looseness=-1

We examine co-occurrences of values with conceptions of race (defined in Section \ref{groupboundaries}) in order to understand how values and racial categories interact within the algorithmic fairness setting (see Figure~\ref{vals-gb} in Appendix for a comprehensive overview of co-occurrences). \looseness=-1
Legal protection is a common way of conceptualizing race across values, reflecting critiques that the ``protected class'' framework is adopted uncritically within the literature. However, other notions of race tend to appear in conjunction with particular values. Specifically, papers that emphasize justice, non-maleficence, and contextuality conceptualize race as a status category more often than as a legal category. Meanwhile, the notion of race as a ``sensitive attribute'', which rarely appears in papers that emphasize justice, is often associated with papers that express performance-related values. Conversely, the social salience conception of race is rare among the papers that emphasize performance but appears frequently in papers that focus on justice and contextuality. These findings highlight that values are differentially expressed through different classification schemas. Consequently, illuminating hidden assumptions about the nature of race can help surface the values embedded in algorithmic fairness research. \looseness=-1

\section{Discussion}

Our results highlight the fact that racial categories do not appear in a homogeneous way in the algorithmic fairness literature. Although legal anti-discrimination frameworks---and, in particular, the U.S. notion of protected classes---appear frequently in the literature, they have not produced a consensus view of race among algorithmic fairness researchers. While prior research has emphasized the algorithmic fairness community's over-reliance on protected classes \cite{benthall19, hanna20}, legal frameworks are only part of the story. In particular, the influence of academic computer science appears in numerous ways. Although algorithmic fairness brings together researchers from many fields, it has important ties to computer science and machine learning communities. FAccT was originally introduced as a workshop at the Conference on Neural Information Processing Systems, a prominent machine learning conference. Moreover, FAccT has been affiliated with the Association for Computing Machinery since 2019. We argue that this context has shaped the use of racial categories within algorithmic fairness frameworks in important ways.
\looseness=-1

While the algorithmic fairness community and FAccT have attempted to orient themselves around ethical principles, they are shaped by disciplinary values from within computer science. In particular, values of performance and generalizability appear often within the annotated papers. Race is frequently abstracted within algorithmic fairness frameworks so that these frameworks can be generalized to other settings. The term ``sensitive attributes'' appears with little attention to the meaning of sensitivity outside of its original context in privacy, rendering this term similarly abstracted from reality. Finally, technical considerations shape racial categories in important ways. Algorithmic limitations and performance concerns drive researchers toward racial schemas that involve fewer categories, often leading to binary operationalizations of race.
\looseness=-1

 The multiracial case provides a clear illustration of both these findings and their consequences. In particular, the multiracial category is typically absent from analysis. This is in keeping with the tendency in computer science toward simple, often binary, classification schemas, as well as the tendency away from small population sizes. Moreover, when multiracial groups are mentioned, they are treated inconsistently, demonstrating the significant flexibility left in the hands of the researcher. This setting also highlights a persistent lack of justifications—or even explanation—around the adoption of racial categories. Despite the fact that multiracial people can be classified in a number of ways under most schemas, these decisions are rarely stated. The history of multiracial statistics highlights the potential to manipulate this flexibility and obscurity toward a number of political goals. \looseness=-1
 
As Jacobs and Wallach \cite{jacobs21} argue, decisions about how to operationalize social constructs such as race are not merely an academic concern but have real, fairness-related consequences. They advocate for making these operationalization decisions explicit in order to make assumptions visible and testable in the name of transparency, accountability, and contestability. We argue that algorithmic fairness researchers should prioritize these visibility practices for defining race, a key construct in the literature that remains underexamined and inconsistently applied yet central to many of the harms the field purports to address. Meanwhile, details about operationalizations of fairness are often explicitly stated within the FAccT literature. Yet these fairness definitions rest on formalizations of social categories (including racial categories) that are rarely detailed or justified and threaten to undermine the project of algorithmic fairness altogether. \looseness=-1

The history of state race-making and racial statistics outlined in Section \ref{racemaking} reveals that racial categories are susceptible to manipulation and have been weaponized to advance state goals. As the work of categorization falls into the hands of algorithmic fairness practitioners, care must be given to ensure that both old and new avenues for manipulation are addressed. The uncritical adoption of existing legal frameworks may reproduce longstanding power relations enacted by the state. On the other hand, the flexibility of racial categories can be leveraged toward other interests, for example, obscuring discrimination within a system by choosing a categorization schema that shows parity between the defined groups. \looseness=-1

The government context offers key lessons for the algorithmic fairness community. Historically, state racialization has been driven by institutional factors within government including evidentiary standards, record-keeping requirements, and incentives \cite{brown20}. We argue that institutions within the algorithmic fairness community will determine how racial categories are instantiated beyond the government context by creating new evidentiary standards, record-keeping requirements, and incentives. From the nascent algorithmic fairness community, Institutions including publishing venues, auditing bodies, and regulatory authorities will determine how race is conceptualized. Thus far, publishing standards and incentives, the persistence of proprietary data, and legal compliance incentives have given shape to a regime under which racial categories are adopted inconsistently and with little expectation of justification. This allows racial categories to be constructed in relative obscurity toward any number of ends, from finding significant scientific results to green-lighting corporate projects. Such flexibility and obscurity merit particular attention in light of recent concerns about corporate capture within FAccT stemming from institutional factors like funding and proprietary data access \cite{young22}. Based on both the historical influence of institutions and the results of our analysis, we argue that institutional contexts and data access shape the adoption of racial categories. We propose that the algorithmic fairness community must work to produce institutional arrangements that center the redistribution of power and promote the community's intended values of fairness, accountability, and transparency. \looseness=-1

FAccT, its reviewers, funding agencies, and the larger algorithmic fairness community all play an important institutional role. Interventions targeting racial discrimination ought to be assessed not only based on their technical details, but also on whether these interventions are built upon a meaningful and relevant understanding of race. In this paper we identify five types of justifications that are used to motivate the adoption of a racial classification schema: data availability, technical motivations, prior scientific work, epistemic concerns, and contextual relevance. While each of these justifications can provide important details, we propose that every justification should center the contextual relevance of its racial classification schema. This information is key to ensuring that readers can understand and evaluate research, use it in the appropriate context, and assess in whose interest a given racial classification schema was chosen. \looseness=-1

The analysis presented in this paper is limited by several factors. First, we focus on what is written directly within the annotated papers. Some authors may choose not to include information about their decisions or include this information in supplemental materials due to space constraints. Because our analysis foregrounds the importance of visibility practices, we argue that the focus on what is included in the body of each paper is justified. However, interviews may elicit a different understanding of how researchers conceptualize race and greater insight into how racial schemas operate as a cognitive process. A second limitation comes from the decision to focus on academic work within a specific conference. This paper does not capture algorithmic fairness research published at other venues (for example, at traditional computer science conferences). However, we show that even when ethical values are explicitly prioritized, as in the case of FAccT, disciplinary values from computer science remain influential. By focusing on FAccT we highlight its role as a key location for institutional change, but this paper does not engage with the significant algorithmic fairness work that occurs outside of academia. Future work should examine how racial categories are implemented in industry and government settings as these are important sites of practice. Finally, this paper covers only the initial years of FAccT from 2018 to 2020. Since then, FAccT has matured and begun to cite a larger canon, including humanistic work and work previously published at FAccT itself. Additionally, the conference has adopted an increasingly reflexive and self-critical orientation, as exemplified by the existing critiques of racial categories in the algorithmic fairness literature \cite{benthall19, hanna20}. While our work focused primarily on the external influences and foundations that the FAccT community drew from in the early years of the conference, current practices merit ongoing attention and reflection.
\looseness=-1

\section{Conclusion}
Important critiques of algorithmic fairness have highlighted the field's failure to account for the complex, socially situated, and political nature of racial categories \cite{benthall19,hanna20, kasirzadeh21, lu22}. We build on this work by examining how algorithmic fairness researchers use racial categories in practice, how they justify these decisions, and the values underlying their choices. Through a systematic qualitative analysis of the FAccT literature, we show that racial categories are inconsistently applied throughout the algorithmic fairness literature, with little justification or explanation. Despite recourse to the language of ``protected classes'' and the state's historical role in racial classification, we find abstract and binary racial schemas are commonly adopted while government racial schemas remain absent. We argue that this points to the importance of computer science, and its values of performance and generalizability, in shaping the field of algorithmic fairness. We also discuss the need for institutional reforms that center visibility practices and careful operationalizations of race. By highlighting the role of values and institutional factors in shaping racial categories, we hope that this work can enable the algorithmic fairness community to re-examine its practices around racial classification in order to align the field's interventions with its values.

\begin{acks}
We are grateful to Matthew Bui, Dallas Card, Meera Desai, Jared Katzman, Lu Xian, and three anonymous FAccT reviewers for helpful comments on earlier versions of this paper. 
\end{acks}

\bibliographystyle{ACM-Reference-Format}
\bibliography{APJ2023}

\onecolumn
\appendix
\section{Supplemental Tables and Figures}
The following tables and figures present counts of various concepts discussed in the results section or co-occurences of these concepts. All counts refer to the number of annotated papers in which each concept appears.

\begin{table}[h]
     \caption{Number of annotated papers that formalize race according to each type of categorization schema. The ``Other'' category encompasses a variety of schemas with more than two race categories. Some papers adopt multiple schemas so the counts do not add up to 60. If a paper includes an abstract formalization, but later uses specific racial categories, it is not counted in the ``Abstract Only'' category. }
    \begin{tabular}{ll}
    \toprule 
    \textbf{Race Category Type} & \textbf{Frequency} \\
    \midrule
    Abstract Only         & 12        \\
    Black/White          & 11        \\
    White/Non-White          & 3         \\
    Black/Non-Black          & 2         \\
    Skin Color          & 2         \\
    Other          & 8         \\
    N/A         & 24        \\
    \bottomrule
    \end{tabular}
    \label{table:race_category_types}
\end{table}

\begin{table}[h]
    \caption{Number of annotated papers that include each type of justification. Justifications that raise ``Epistemic Concerns'' include references to consistency, reliability, objectivity, and/or precision.  Some papers use multiple justifications so the counts do not add up to 60.}
    \begin{tabular}{ll}
    \toprule
    \textbf{Justification Type}              & \textbf{Frequency} \\
    \midrule 
    None (specific race categories) & 15 \\               
    Relevance & 5 \\ 
    Prior Work & 4 \\
    Data Availability & 4 \\ 
    Technical & 4 \\ 
    Epistemic Concerns & 3 \\ \\
    None (abstract or no race categories) & 32 \\ 
    \bottomrule
    \end{tabular}
    \label{table:justification_types}
\end{table}

\begin{table}[h]
\caption{Boundaries between racial groups are constructed in terms of numerous axes of difference. This table summarizes the number of annotated papers in which each type of boundary appears and includes illustrative quotes of how these boundaries are expressed in the papers. Most papers conceptualize race in multiple ways so the counts do not add up to 60.}
\begin{tabular}{lll}
\toprule
\textbf{Group Boundary}                   & \textbf{Frequency} & \textbf{Examples}                                                                                                                                                                   \\
\midrule
Legal Protection       & 27    & ``members of protected classes'' \\ 
& & ``protected demographic groups'' \\
& & ``characteristics legally protected against discrimination''                                          \\

Status                 & 25    & ``historically marginalized groups'' \\ 
& & ``advantaged and disadvantaged groups'' \\ 
& & ``members of a dominant group'' \\ 
& & ``status categories'' \\ 
& & ``privileged group''                  \\

Minoritization         & 17    & ``majority group'' \\ 
& & ``minority races'' \\ 
& & ``racial and ethnic minorities'' \\

Sensitivity & 17    & ``sensitive attribute'' \\ 
& & ``sensitive feature'' \\ 
& & ``sensitive group membership''                                                                                             \\

Social Salience        & 10    & ``socially salient group'' \\ 
& & ``groups relevant to fairness analysis'' \\ 
& & ``socially significant [...] groups''                                                                \\

Label                  & 7     & ``people labeled `White' and `Hispanic'\thinspace'' \\ 
& & ``defendants whose race is recorded as either African-American or Caucasian'' \\ 
& & ``racial labels''                                \\

Observable Differences & 6     & ``darker-skinned'' \\ 
& & ``Afrocentric features'' \\ 
& & ``black-associated names'' \\ 
& & ``coder’s[\textit{sic}] [...] relied on photos or language in the absence of self-identification'' \\

Identity               & 4     & ``self-reported race/ethnicity'' \\ 
& & ``what categories a person individually identifies with'' \\ 
& & ``consent to and define their own categorization''                            \\

Origin                 & 2     & ``people of African descent'' \\ 
& & ``European demographics'' \\ 
\bottomrule
\end{tabular}
\label{table:noname_types}
\end{table}

\begin{figure}[h]
  \centering
  \includegraphics[width=\linewidth]{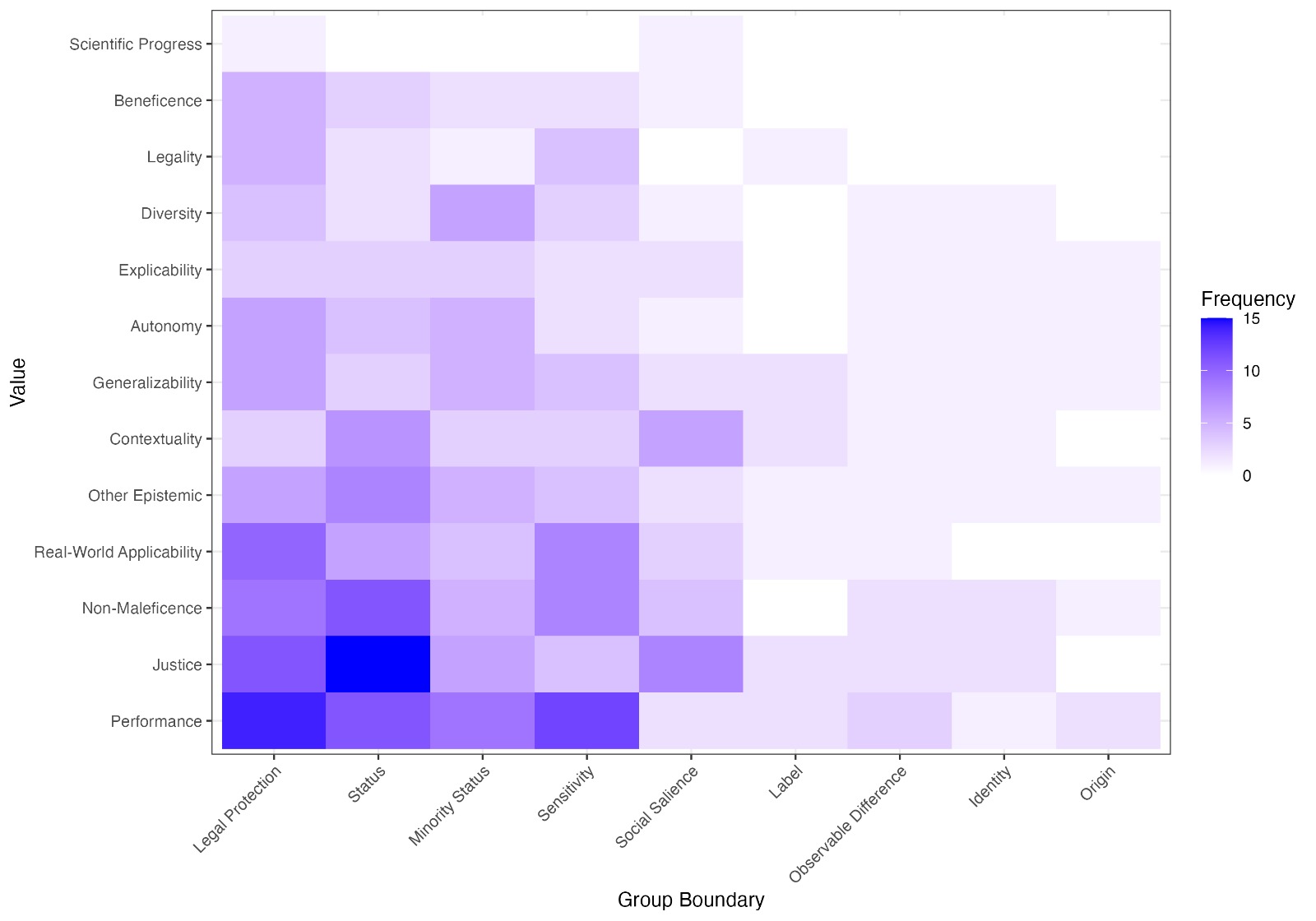}
  \caption{A heat-map showing co-occurences of values with conceptualizations of boundaries between racial groups. Darker squares represent a higher number of papers in which a given pair co-occur. Terms on the axes are ordered bottom-to-top and left-to-right from highest frequency to lowest frequency.}
  \label{vals-gb}
  \Description{A heat-map with values on the x-axis and group boundaries on the y-axis.}
\end{figure}

\end{document}